\documentclass[reprint,prb]{revtex4-2}
\usepackage{tikz}
\usepackage{circuitikz}
\usepackage{pst-solides3d}

\usepackage[utf8]{inputenc}
\usepackage{setspace,amsmath,amsfonts,amssymb,esint}
\usepackage{hyperref}
\usepackage{mathptmx}
\usepackage{mathrsfs}

\newcommand{\re}[1]{\mbox{Re}\left\{#1\right\}}

\newcommand{\n}{\hat{\mathbf{n}}}
\newcommand{\tv}{\hat{\mathbf{t}}}

\newcommand{\rb}{\mathbf{r}}
\newcommand{\rt}{\tilde{\mathbf{r}}}
\newcommand{\rbt}{\tilde{\mathbf{r}}}
\newcommand{\rp}{\left( \tilde{\mathbf{r}} \right)}
\newcommand{\rpp}{\left( \tilde{\mathbf{r}}' \right)}

\newcommand{\tOmega}{ \tilde{\Omega} }

\newcommand{\tGamma}{ \tilde{\Gamma} }
\newcommand{\dV}{d\tilde{V}}

\newcommand{\dl}{d\tilde{l}}

\makeatletter
\newsavebox{\@brx}
\newcommand{\llangle}[1][]{\savebox{\@brx}{\(\m@th{#1\langle}\)}%
  \mathopen{\copy\@brx\kern-0.5\wd\@brx\usebox{\@brx}}}
\newcommand{\rrangle}[1][]{\savebox{\@brx}{\(\m@th{#1\rangle}\)}%
  \mathclose{\copy\@brx\kern-0.5\wd\@brx\usebox{\@brx}}}
\newcommand{\Jto}{{\bf j}}
\newcommand{\gammato}[1]{\kappa^{\perp}_{#1}}

\colorlet{LightGray}{gray!10}
\makeatother

\newcommand{\s}[1]{\underline{\text{#1}}}
\newcommand{\ds}[1]{\underline{\underline{\text{#1}}}}

\begin{document}

\begin{abstract}
The electromagnetic scattering from interconnections of high-permittivity dielectric thin wires with sizes smaller than (or almost equal to) the operating wavelength is investigated. A simple lumped element model for the polarization current intensities induced in the wires is proposed. The circuit elements are capacitances and inductances between the wires. An analytical expression for the induced polarization currents in terms of the magneto-quasistatic current modes is obtained. The connection between the spectral properties of the loop inductance matrix and the network's resonances is established. The number of the allowed current modes and resonances is deduced from the topology of the circuit's digraph. The coupling to radiation is also included, and the radiative frequency shifts and the quality factors are derived. 
The introduced concept and methods may find applications both at the microwaves and in nanophotonics.
\end{abstract}

\title{Electromagnetic Scattering by Networks of High-Permittivity Thin Wires}

\author{Carlo Forestiere}
\affiliation{ Department of Electrical Engineering and Information Technology, Universit\`{a} degli Studi di Napoli Federico II, via Claudio 21,
 Napoli, 80125, Italy}
\author{Giovanni Miano}
\affiliation{ Department of Electrical Engineering and Information Technology, Universit\`{a} degli Studi di Napoli Federico II, via Claudio 21,
 Napoli, 80125, Italy}
 \author{Bruno Miranda}
\affiliation{ Department of Electrical Engineering and Information Technology, Universit\`{a} degli Studi di Napoli Federico II, via Claudio 21,
 Napoli, 80125, Italy}
\affiliation{Institute of Applied Sciences and Intelligent Systems - Unit of Naples, National Research Council, via P. Castellino 111, Naples, 80131 Italy.}

\maketitle
\label{sec:SDA}

\section{Introduction}


High-permittivity dielectric objects \cite{richtmyer_dielectric_1939} are currently intensively studied both at the microwave and in the visible, promising diverse applications including microwave antennas \cite{long_resonant_1983}, nanoscale biosensors \cite{yavas_--chip_2017}, wireless mid-range energy transfer \cite{karalis_efficient_2008}, non-linear optics \cite{koshelev_subwavelength_2020}, and metamaterials \cite{holloway_double_2003}. Resonances of small high-permittivity dielectric objects arise from the interplay between the polarization energy of the dielectric and the energy stored in the magnetic field. They can be described within the magneto-quasistatic approximation of the Maxwell's equation \cite{forestiere_magnetoquasistatic_2020}, extended by radiative corrections \cite{forestiere_resonance_2020}. In the magneto-quasistatic approximation the induced polarization density field is solenoidal in the object and its normal component to the surface of the object is equal to zero \cite{bladel_resonances_1975}. These features stimulate the search of relevant geometries for which high-permittivity dielectric resonators behave as lumped networks when their size is smaller or comparable to the operating wavelength.
A possible candidate is an arbitrary interconnection of high-permittivity dielectric thin wires, that we denote in the following as {\it high-permittivity dielectric network}. A wire is {\it thin} if the linear dimension of its cross section is much smaller than its length. 

In this paper, we propose a lumped element model to describe the electromagnetic scattering from high-permittivity networks, when the size  is smaller than (or at almost equal to) the operating wavelength. 
We develop the model in the framework of Maxwell's equation, by expanding the polarization currents in terms of magneto-quasistatic modes, and by taking into account the radiation perturbatively \cite{forestiere_magnetoquasistatic_2020,forestiere_resonance_2020}. We carried out the lumped element model in terms of capacitances,  self- and mutual inductances between the wires. This simplified model leads to the analytical prediction of: the polarization currents induced in the dielectric network by an external electromagnetic field; the magneto-quasistatic current modes and resonances of the dielectric network; their radiative frequency shifts and quality factors. Specifically, we found that the spectral properties of the circuit's loop inductance matrix determine the electromagnetic scattering resonances and modes of the dielectric network. Thus, instead of solving Maxwell's equation, that would be unpractical for intricate connections of multiple branches, we import concepts and formulas that have been produced by scientists and engineers working on electric inductive network \cite{paul_inductance_2011}, in particular the analytic formulas produced in the first twenty years of the XX century and summarized in the manuscripts of Rosa \cite{rosa_self_1908}, Grover \cite{grover_inductance_nodate}, and Weber \cite{weber_electromagnetic_1950}. These concepts are now applied to the polarization current densities, rather than to the electric currents, consistently with the framework proposed by Engheta and co-workers in Refs. \cite{engheta_circuit_2005,alu_optical_2007,alu_parallel_2007,salandrino_parallel_2007}. 
In particular, one of the main strengths of the proposed model lies in its simplicity: when the network consists of only a few loops, the calculation of the magneto-quasistatic current modes and the associated resonances can be carried out with just paper and pencil, together with their radiative frequency shifts and quality factors. It also enucleates the connection between the current modes, resonances, radiative corrections, and the topology of the underlying graph. It may also help the comprehension of lasing in complex photonic graphs and networks \cite{lepri_complex_2017,gaio_nanophotonic_2019,lubatsch_self-consistent_2019,massaro_heterogeneous_2021}.

The paper is organized as follows. After a brief summary of the properties of the magneto-quasistatic current modes, resonances and radiative corrections, we introduce - through examples of increasing complexity - the steps required to assembly a lumped-element model for complex high-permittivity dielectric networks. First, in Sec. \ref{sec:IsolatedLoop} we consider an isolated high-permittivity loop, and we link its self-inductance with the resonance frequency and the radiative quality factor. Then, in Sec. \ref{sec:TwoLoops}, we extend the lumped element model to two interacting high-permittivity loops. Eventually, we consider in Sec. \ref{sec:circuit} an arbitrary interconnection of high permittivity thin wires.  
\section{Electromagnetic scattering from a high-permittivity dielectric object}

In the linear regime, resonant electromagnetic scattering from nonmagnetic and small objects, occurs according to two different mechanisms (e.g., \cite{forestiere_magnetoquasistatic_2020,forestiere_magnetoquasistatic_2020}, \cite{forestiere_resonance_2020}). If the real part of the permittivity is negative (e.g., as in metals) resonances arise from the interplay between the electric field energy of the electro-quasistatic current modes and the polarization energy. Instead, if the real part of the permittivity is positive and very high (e.g., as in dielectrics) resonances arise from the interplay between the magnetic field energy of the magneto-quasistatic current modes and the polarization energy. The quasistatic current density modes are solutions of the source-free Maxwell equations in the quasistatic limits. In particular, the magneto-quasistatic current density modes are solenoidal in the object and their normal component to the boundary of the object is equal to zero. In the following, we briefly resume the principal features of the magneto-quasistatic current modes together with the radiative corrections (for details see in \cite{forestiere_magnetoquasistatic_2020,forestiere_magnetoquasistatic_2020}, \cite{forestiere_resonance_2020}). 

Let us consider an isotropic and homogeneous dielectric surrounded by vacuum and occupying a volume $\Omega$; $\partial \Omega$ is the boundary of $\Omega$ with outward-pointing normal $\n$. The object is illuminated by a time harmonic electromagnetic field incoming from infinity $\text{Re} \left\{ \mathbf{E}_{ext} \left( \mathbf{r} \right) e^{i \omega t } \right\}$, where $\omega$ is the angular frequency. 
We denote with $\varepsilon_R$ the relative permittivity of the object, with $\chi = \varepsilon_R - 1$ the susceptibility, and with $\varepsilon_0$ the vacuum permittivity. 

We now introduce the dimensionless size parameter of the object $ x = \omega/\omega_c$ where $\omega_c= c_0/l_c$, $l_c$ is the radius of the minimum sphere circumscribing the dielectric object, and $c_0$ is the light velocity in vacuum; $x$ is equal to the ratio between $l_c$ and the operating wavelength in vacuum. 

\subsection{Magneto-quasistatic approximation}
\label{sec:Fundamental}

In objects with high relative permittivity, $\epsilon_R\gg1$, and with sizes much smaller than the operating wavelength, i.e., $x \ll 1$, the contribution of the magneto-quasistatic current density modes $\left\{\Jto_h\right\}$ to the induced polarization current density field is given by \cite{forestiere_magnetoquasistatic_2020,forestiere_magnetoquasistatic_2020}:
\begin{equation}
   \label{eq:expansion}
   \mathbf{J}_p \left( \rt \right) = i \omega \chi \varepsilon_0 \sum_h \frac{1} {1 - x^2 \chi/\gammato{h} }\langle \Jto_h, \mathbf{E}_{ext} \rangle \,  \Jto_h \rp
\end{equation}
where $\Jto_h \rp$ are solutions of the eigenvalue problem
\begin{equation}
\frac{1}{4\pi} \int_{\tOmega} \frac{ \Jto_h \rpp }{\left| \rt- \rt' \right|} \dV' = \frac{1}{\gammato{h}} \Jto_h\rp, \quad \forall \rt \in \tOmega
\label{eq:MQSproblem}
\end{equation}
with
\begin{equation}
 \Jto_h \rp \cdot \n \rp=0 \qquad \forall \rt \in \partial \tOmega,
 \label{eq:boundary}
\end{equation}
and $\gammato{h}$ is the eigenvalue associated to the eigenfunction $\Jto_h\rp$; $\langle \mathbf{A}, \mathbf{B} \rangle = \int_{\tOmega} \mathbf{A} \cdot \mathbf{B} \, dV$ is the standard inner product. In Equations \ref{eq:expansion}-\ref{eq:boundary} the position vector $\rb$ have been normalized by $l_c$, i.e., $\rt = \rb /l_c$, and $\tOmega$ is the corresponding scaled domain and $\partial\tOmega$ its boundary. 

Apart from the factor $\mu_0$, the integral operator on the left hand side of equation \ref{eq:MQSproblem} gives the static magnetic vector potential in the Coulomb gauge as function of the current density field (here we call it the magneto-quasistatic integral operator). Equation \ref{eq:MQSproblem} holds in weak form in the functional space equipped with the inner product $\langle \mathbf{A}, \mathbf{B} \rangle$, and constituted by the vector fields that are solenoidal in $\tOmega$ and have zero normal component to $\partial \tOmega$. The spectrum of the magneto-quasistatic integral operator with the boundary conditions \ref{eq:boundary} is discrete $\left\{ \gammato{h} \right\}_{h \in \mathbb{N}}$, and the eigenvalues are real and positive. The current density modes $\left\{ \Jto_h \right\}_{h \in \mathbb{N}}$ are solenoidal in $\tOmega$, have zero normal component to $\partial \tOmega$, and are orthonormal according to the scalar product $\langle \mathbf{A}, \mathbf{B} \rangle$. The modes are ordered in such a way $\gammato{0}<\gammato{1}<\gammato{2}<...$ .
The expression \ref{eq:expansion} has been obtained by solving the full wave electromagnetic scattering problem in the quasistatic limit (for detail see in \cite{forestiere_magnetoquasistatic_2020,forestiere_magnetoquasistatic_2020}). Here we have disregarded the contribution of the electro-quasistatic current density modes because they give nonresonant terms in the electromagnetic scattering from a high-index dielectric.

The resonance frequency $\omega_h^\perp$ of the $h$-th mode $\Jto_h$ is obtained by maximizing its amplitude in the expression \ref{eq:expansion}. It is solution of the equation
\begin{equation}
\left(\frac{\omega_h^\perp}{\omega_c}\right)^2= \frac{ \gammato{h}}{\re{\chi(\omega_h^\perp)}}.
  \label{eq:MQSxres0}
\end{equation}

 \subsection{Radiative corrections}
\label{sec:Fundamental}

When the dimensionless size parameter of the object becomes almost equal to one, $x\simeq1$, the set of modes $\{\Jto_h\}$ still approximates well the modes of the objects, nevertheless the magneto-quasistatic approximation cannot provide their radiative shift and quality factors.  The results obtained by magneto-quasistatic approximation have to be supplemented by radiative corrections \cite{forestiere_resonance_2020}. Indeed, expression \ref{eq:expansion} is still a valid approximation for the induced polarization current density when $x\simeq1$ as long as $\gammato{h}$ is replaced with
\begin{equation}
\kappa_h = \gammato{h} + \kappa^{\left(1\right)}_h x +  \kappa^{\left(2\right)}_h x^2+ \ldots 
\end{equation}
where $\kappa^{\left(n\right)}_h$ is the $n$-th coefficient in the expansion of $\kappa_h$ in a power series of $x$ (for details see in \cite{forestiere_resonance_2020}). For any object shape it results $\kappa^{\left(1\right)}_h=0$. Taking  into account the second order (real) correction $\kappa^{\left(2\right)}_h$, we obtain that the resonance frequency is solution of the equation
\begin{equation}
\left(\frac{\omega_h}{\omega_c}\right)^2 =\frac{\gammato{h}}{\re{\chi(\omega_h)}- {\kappa_{h}^{\left(2\right)}}}. 
  \label{eq:MQSxres2}
\end{equation}
The coefficient $\kappa^{\left(2\right)}_h$ is negative, therefore the second order radiative correction introduces a negative frequency shift with respect to the quasistatic resonance given by \ref{eq:MQSxres0}. The first non-vanishing imaginary correction of order $n_i \ge 3$, has a positive imaginary part that we denote with $\kappa^{\left(n_i\right)}_h$. It determines the radiative broadening of the mode, related to the inverse of the quality factor, that is given by \cite{forestiere_resonance_2020}
\begin{equation}
    Q_h = \frac{\kappa^{\perp}_h}{\kappa^{\left(n_i\right)}_h} \left( \frac{1}{x_h}\right)^{n_i}
    \label{eq:RadiativeQ},
\end{equation}
where $x_h$ is the size parameter evaluated at the resonance frequency given by \ref{eq:MQSxres2}. The order of the first non-vanishing imaginary correction $n_i$ returns the multipolar scattering order \cite{forestiere_resonance_2020}. For instance, the modes with $n_i=3$ exhibit a non-vanishing magnetic dipole moment, the ones with $n_i=5$ have either magnetic quadrupole or $\text{P}_{\text{E}2}$ moment \cite{bladel_hierarchy_1988} (also called toroidal dipole)  different from zero, etc.. The general expressions for the radiation corrections of magneto-quasistatic current density modes are given in Ref. \cite{forestiere_resonance_2020}.

\section{Isolated high-permittivity thin wire loop}
\label{sec:IsolatedLoop}

We now consider a high-permittivity thin wire dielectric loop with uniform cross section $\Sigma$  (the transverse linear dimensions are much smaller compared to the length). The wire axis is represented by the closed curve $\Gamma$ with tangent unit vector $\hat{ \bf t}$.  With abuse of notation  we also indicate with $\Sigma$ the cross-sectional area, with $\Gamma$ the loop length, and with $\Omega$ the loop volume, which is given by $\Omega=\Gamma\times\Sigma$.

\subsection{Fundamental magneto-quasistatic current mode}
\label{sec:Fundamental}
In a loop with small cross section, the lowest eigenvalue $\gammato{0}$ is well separated from the remaining eigenvalues, as shown in \cite{forestiere_magnetoquasistatic_2020} for a torus of finite cross section. This separation increases as the wire cross section reduces. This fact suggests that for $x$ smaller than (or almost comparable to) one in a thin wire dielectric loop only the mode $\Jto_0 \rp$ associated to the smallest eigenvalue $\gammato{0}$ (fundamental magneto-quasistatic mode) is excited. 
Therefore, in this case the expression of the induced polarization current density \ref{eq:expansion} reduces to
\begin{equation}
   \label{eq:expansion1}
   \mathbf{J}_p \left( \rt \right) = i \omega \chi \varepsilon_0 \frac{1}{1 - x^2 \chi/\gammato{0} }\langle \Jto_0, \mathbf{E}_{ext} \rangle \\  \Jto_0 \rp.
\end{equation}
The fundamental mode $\Jto_0 \rp$ is directed along the wire axis $\hat{\bf t}$, and its module is uniform in $\tOmega$ (as shown in \cite{forestiere_magnetoquasistatic_2020} for a torus of finite cross section),
\begin{equation}
\label{eq:mode} 
    \Jto_0 \rp= \left\{
    \begin{array}{cl}
         \hat{\bf t}\rp/\sqrt{\tOmega} & \text{in} \;\tOmega,  \\
          \bf 0 & \text{otherwise.}
    \end{array}
    \right.
\end{equation}
The mode is normalized in such a way $\langle \Jto_0, \Jto_0 \rangle=1$. The corresponding eigenvalue $\gammato{0}$ is given by Eq. \ref{eq:MQSproblem}, i.e.
\begin{equation}
\frac{\tOmega}{\gammato{0}}= \frac{1}{4\pi}\int_{\tOmega}\int_{\tOmega}  \frac{ \hat{\bf t} \rp\cdot\hat{\bf t} \rpp }{\left| \rt- \rt' \right|} \dV' \dV,
   \label{eq:Kappa2} 
\end{equation}
where $\tOmega$ on the left hand side denotes the scaled volume of the object ($\tOmega=\Gamma\times\Sigma/l_c^3$). Apart from the factor $(l_c^5/\Sigma^2)\mu_0$, the integral of the right hand side is equal to the self-inductance of the loop $L$, therefore
\begin{equation}
\gammato{0}=\frac{l_c\Gamma}{\Sigma} \frac{l_c\mu_0}{L}.
   \label{eq:KappaL} 
\end{equation}

\subsection{Lumped element model for a single loop}
\label{sec:circuitmodel}
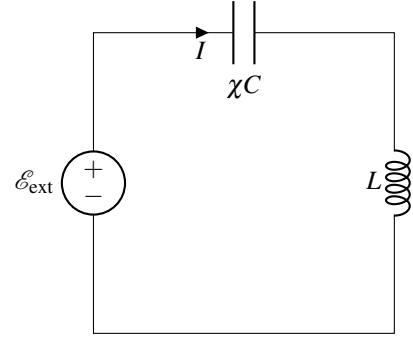
\begin{figure}
\centering
\begin{circuitikz}
\draw (0,0) to [C,l_=$\chi C$,i>_=$I$] (2,0) -| (3,-1) to [L,l_=$L$] (3,-3) |- (2,-4) to (0,-4) -| (-1,-3) to [american voltage source,invert,l=$\mathcal{E}_{\text{ext}}$] (-1,-1) |- (0,0);
\end{circuitikz}
    \caption{Lumped element circuit for a high-permittivity loop}
    \label{fig:CircuitLoop}
\end{figure}

The expression \ref{eq:expansion1} has a very simple physical explanation.  The polarization current density field induced in the loop is given by
\begin{equation}
\label{eq:constitutive}
    \mathbf{J}_p = i \omega \varepsilon_0 \chi \left( {\bf E}  + {\bf E}_{ext} \right) 
\end{equation}
where ${\bf E}$ is the induced electric field. By performing the line integral along $\Gamma$ on both sides of equation \ref{eq:constitutive} we obtain
\begin{equation}
  \frac{1}{\chi} \frac{1}{i\omega  C } {I} =   \mathcal{E}_{ext} + \mathcal{E}
  \label{eq:LKT}
\end{equation}
where $I$ is the intensity of the induced polarization current in the wire, 
\begin{equation}
    C= \varepsilon_0\frac{\Sigma}{\Gamma},
    \label{eq:capacity}
\end{equation}
$\mathcal{E}$ is the induced voltage along the loop
\begin{equation}
    \mathcal{E}  = \oint_\Gamma {\bf E} \cdot \tv \, dl,
\end{equation}
and $\mathcal{E}_{ext}$ is the applied voltage
\begin{equation}
    \mathcal{E}_{ext} = \oint_\Gamma {\bf E}_{ext}\cdot \tv dl.
\end{equation}
On the other hand, from the Neumann-Faraday law we obtain:
\begin{equation}
     \mathcal{E}  =- i \omega L I
     \label{eq:Voltage}
\end{equation}
where 
$L$ is the self-inductance of the loop. Using Eq. \ref{eq:Voltage} in Eq. \ref{eq:LKT} we have:
\begin{equation}
   \left( i \omega L+\frac{1}{\chi} \frac{1}{i\omega  C } \right) I = \mathcal{E}_{ext},
\end{equation}
which is the equation governing the equivalent circuit in Fig. \ref{fig:CircuitLoop}. Since
\begin{equation}
  \frac{1}{\omega_c ^2 \gammato{0}}=LC,
   \label{eq:Loneloop}
\end{equation}
the solution of equation \ref{eq:Loop1} is expressed as 
\begin{equation}
  I =i\omega C\chi \frac{1}{1 - \left(\omega/\omega_c \right)^2 \chi/\gammato{0}} \mathcal{E}_{ext} .
   \label{eq:Loop1}
\end{equation}
This expression coincides with the expression of the polarization current intensity obtained from \ref{eq:expansion1}. 

The quasi-static resonance frequency, solution of equation \ref{eq:MQSxres0}, is the value of $\omega$ for which  $\left|{1 - \left(\omega/\omega_c \right)^2 \chi/\gammato{0}}\right|$ is minimum.
\subsection{Radiative corrections}
In this case the radiative corrections $\kappa^{\left(2\right)}_h$ and $\kappa^{\left(3\right)}_h$ have the following expressions \cite{forestiere_resonance_2020}:
\begin{align}
    \label{eq:Kappa2b}
    \kappa^{\left(2\right)}_0 &=  \frac{ \left(\gammato{0}\right)^2}{8\pi} \int_{\tOmega} \int_{\tOmega}  \left| \rbt -\rbt'\right| \, \Jto_0 \rp\cdot\Jto_0 \rpp \dV' \dV, \\
    \kappa^{\left(3\right)}_0 &=  -i \frac{ \left(\gammato{0}\right)^2}{24\pi} \int_{\tOmega} \int_{\tOmega}  \left| \rbt -\rbt'\right|^{2} \, \Jto_0 \rp\cdot \Jto_0 \rpp \dV' \dV
    \label{eq:Kappa_ni} .
\end{align}
By using \ref{eq:mode}, expression \ref{eq:Kappa2b} becomes
\begin{equation}
\kappa_0^{\left( 2 \right)} =  -2\pi\frac{ \Sigma}{\Gamma ^2} { \left(\gammato{0}\right)^2}
\Delta
\end{equation}
where
\begin{equation}
    \Delta = -\frac{1}{8\pi} \oint_{\tGamma} \oint_{\tGamma} \tv \rp \cdot \tv \rpp {\left| \rbt -\rbt'\right|}  \dl' \dl > 0;
\end{equation}
we have chosen $l_c=\Gamma/2\pi$. By using \ref{eq:mode} expression \ref{eq:Kappa_ni} becomes:
\begin{equation}
   \kappa_0^{\left( 3 \right)}  = i \left(\gammato{0} \right)^2  \frac{1}{6\pi} \left\| \mathbf{P}_{\text{M}} \right\|^2,
   \label{eq:Kappa3}
\end{equation}
where $\textbf{P}_{\text{M}}$ is the magnetic dipole moment of the mode in the scaled object,
\begin{equation}
    \textbf{P}_{\text{M}} = \frac{1}{2} \int_{\tOmega} \rbt \times \mathbf{j}_0 \,  \dV.
    \label{eq:DipoleM}
\end{equation}
The radiation quality factor is given by
\begin{equation}
    Q = \frac{6 \pi}{  \kappa_0^\perp \left\| {\bf P}_{\text{M}} \right\|^2} \frac{1}{x_{r}^3},
\end{equation}
where $x_{r}=\omega_{r}/\omega_c$ and $\omega_{r}$ is the resonance frequency; it is solution of equation \ref{eq:MQSxres2} and takes into account the radiative shift.

The expression of the induced polarization current in the loop with the radiative corrections is obtained by substituting $\kappa_0^\perp $ with $ \kappa_0= \gammato{0} + \kappa^{\left(2\right)}_0 x^2+ \kappa^{\left(3\right)}_0 x^3 $ into expression \ref{eq:Loop1}.
\begin{figure}
\centering
\includegraphics[width=\columnwidth]{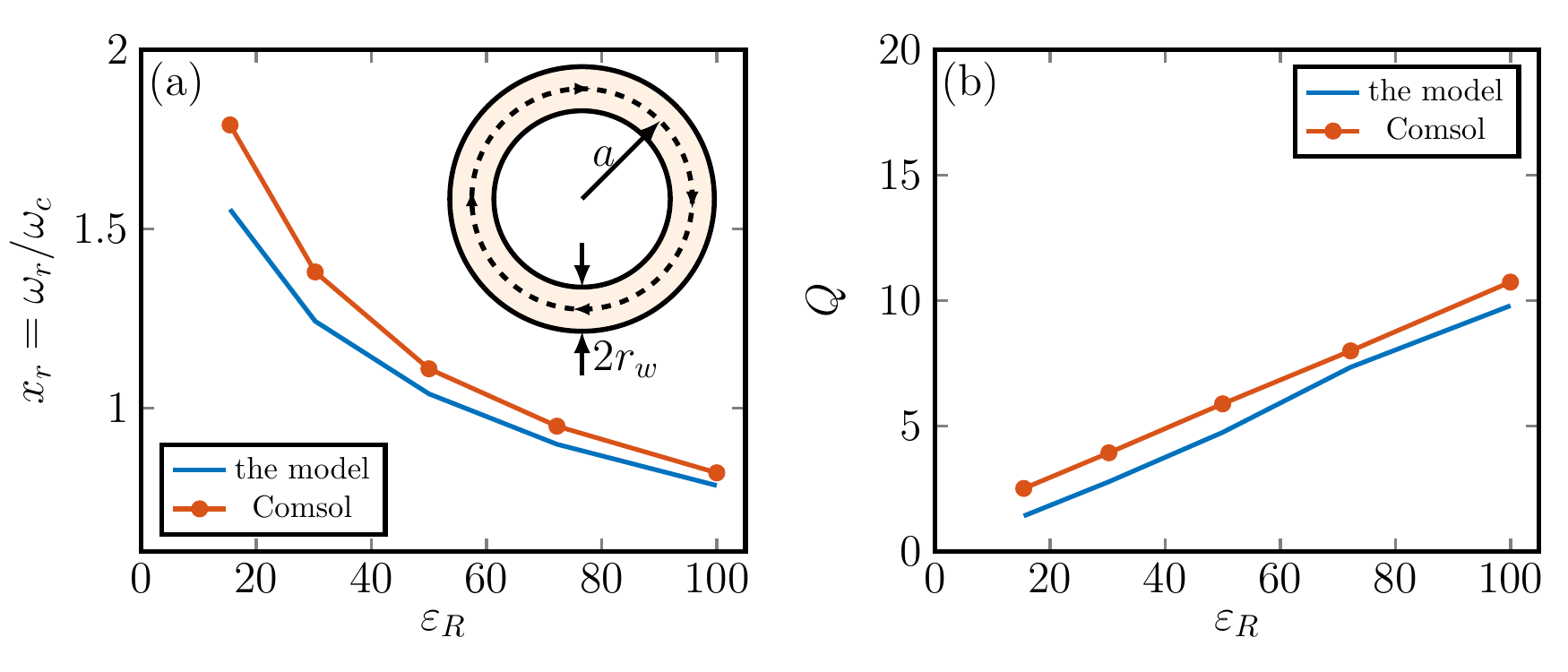}
\caption{Comparison between the results obtained by the proposed model and by Comsol for a circular loop: (a) Resonance frequency normalized to $\omega_c$, $x_{r}=\omega_{r}/\omega_c$, and (b) radiative quality factor of a high-permittivity loop with major radius $a$ and minor radius $r_w=0.1 a$ as a function of the permittivity $\varepsilon_R$. }
\label{fig:CircularLoop}
\end{figure}
\subsection{Circular loop}
To illustrate the use of the formulas derived above, we now consider a circular loop (ring) with circular cross section (torus), minor radius $r_w$ and major radius $a$. Hollow nanodisks have been recently used to tailor the magnetic dipole emission \cite{feng_all-dielectric_2016}. The circular loop was the platform where dielectric resonances have been first proposed \cite{richtmyer_dielectric_1939}, and its modes were studied in Ref.
\cite{verplanken_magnetic-dipole_1979,forestiere_magnetoquasistatic_2020}. A circuit model of the dielectric circular loop has been proposed in Ref. \cite{jelinek_artificial_2009}, and bulk metamaterials made of rings have been also investigated \cite{marques_bulk_2011}.

For a circular loop the expression of $\kappa_0^\perp$ is
\begin{equation}
 \kappa^\perp_0 \, =   \frac{2a^2}{r_w^2} \frac{a\mu_0}{L},
 \label{eq:SelfCircLoop}
\end{equation}
where $L$ is the inductance of the loop \cite{paul_inductance_2011},
\begin{equation}
     L = \mu_{0}\sqrt{a\left(a-r_{\mathrm{w}}\right)}\left[\left(\frac{2}{k}-k\right) K(k)-\frac{2}{k} E(k)\right] + \frac{\mu_0}{4} a,
     \label{eq:SelfCircLoop}
\end{equation}
\begin{equation}
    k=\sqrt{\frac{4 a\left(a-r_{\mathrm{w}}\right)}{\left(2 a-r_{\mathrm{w}}\right)^{2}}},
\end{equation}
$K \left( k \right)$ and $E \left( k \right)$ are the complete elliptic integrals of the first and second kind, where the second term in Eq. \ref{eq:SelfCircLoop} represents the internal inductance of the ring \cite{paul_inductance_2011}. The expression of the second order correction is
\begin{equation}
 \kappa_0^{\left( 2 \right)} =  - \frac{2}{3} \frac{2 a^2}{r_w^2} \left( \frac{a\mu_0}{L} \right)^2.
\end{equation}
Therefore for a non dispersive material the expression of the resonant frequency is:
\begin{equation}
 \omega_{r} =
  \frac{c_0}{a}\left( \sqrt{ \frac{r_w^2}{2a^2} \left(\frac{L}{a\mu_0}\right) \text{Re} \left\{ \chi \right\} + \frac{2}{3}  \left( \frac{a\mu_0}{L} \right)} \right)^{-1}.
  \label{eq:lambdaCircLoop}
\end{equation}
The expression of the radiative quality factor is given by:
\begin{equation}
    Q =  \frac{6}{\pi} \left(\frac{L}{a\mu_0} \right) \frac{1}{x_{r}^3}
    \label{eq:QCircLoop}.
\end{equation}

We have validated formulas \ref{eq:lambdaCircLoop} and \ref{eq:QCircLoop} against Comsol multiphysics (wave optics package) considering a high permittivity solid torus with $r_w =0.1a$. Specifically, in Fig. \ref{fig:CircularLoop}, we compare the resonance position obtained by Eq. \ref{eq:lambdaCircLoop} against the position of the resonance peak of the scattering spectrum computed in Comsol, when the torus is excited by an electric point dipole, located in the equatorial plane of the torus, at a distance $3a$ from the center, and oriented along the toroidal direction.
We found very good agreement for very high permittivities, while the error slightly deteriorates for smaller values of permittivity where the dimension of the object becomes comparable to the resonance wavelength. In Fig. \ref{fig:CircularLoop} (b) we compare the quality factor obtained by Eq. \ref{eq:QCircLoop} against the inverse of the full-width at half maximum of the resonance peak obtained in Comsol. Good agreement is found.

\section{Two interacting high-permittivity loops}
\label{sec:TwoLoops}

In this section, we consider a pair of high-permittivity thin wire loops occupying two disjoint spatial domains $\Omega_1$ and $\Omega_2$ with cross sections $\Sigma_1$ and $\Sigma_2$. The wire axes are represented by the closed curve $\Gamma_1$ and $\Gamma_2$ with tangent unit vector $\hat{ \bf t}_1$ and $\hat{ \bf t}_1$, respectively. The polarization current density in each loop is uniformly distributed across the wire cross-section and directed along its axis; $I_1$ and $I_2$ are, respectively, the polarization current intensities of the two loops.
\subsection{Lumped element model for a loop pair}

\begin{figure}
\centering
\includegraphics[width=\columnwidth]{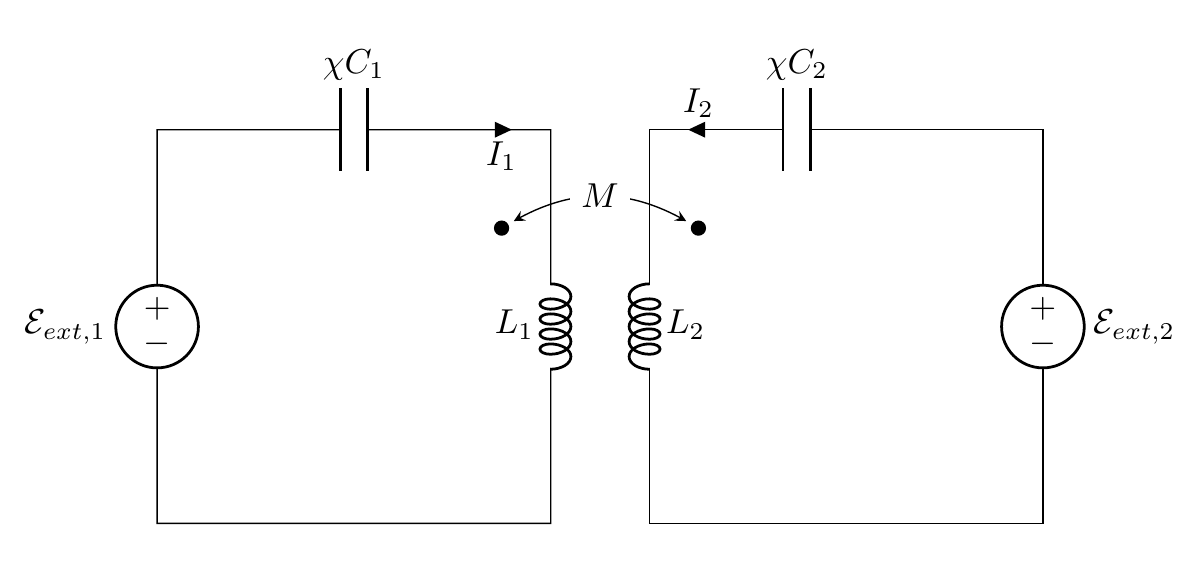}
\caption{Lumped element circuit for two mutually coupled high-permittivity loops.}
\label{fig:Circuit2loops}
\end{figure}

By following the procedure introduced in section \ref{sec:circuitmodel}, we obtain the following system of equations for the intensities of the induced polarization currents (the lumped element circuit is shown in  Fig. \ref{fig:Circuit2loops})
\begin{equation}
  \begin{aligned}
i\omega L_1 I_1  + i\omega M I_2 + \frac{1}{i\chi \omega C_1} I_1&=\mathcal{E}_{ext,1} , \\
 i\omega M I_1  + i\omega L_2 I_2 + \frac{1}{i\chi \omega C_2} I_2&=\mathcal{E}_{ext,2} , \\.  \\
    \end{aligned}
    \label{eq:Loops2external}
\end{equation}
where 
\begin{equation}
    C_q=\varepsilon_0\frac{\Sigma_q}{\Gamma_q} \quad q \in {1,2},
\end{equation} 
$L_1$ and $L_2$ are the self-inductions of the loops, and $M$ is the mutual inductance between the two loops,
\begin{equation}
    {M} = \frac{\mu_0}{4\pi} \oint_{\Gamma_1}\oint_{\Gamma_2} \frac{{{\tv_1 \cdot{\tv_2}}}}{\left| \rb - \rb' \right|} dl_1 dl_2.
    \label{eq:Neumann}
\end{equation}
The modes of the loop pair are solutions of the generalized eigenvalue problem
\begin{equation}
    \ds{L} \, \s{u} = \frac{1}{\kappa^\perp}\frac{1}{\omega_c^2}\ds{C}^{-1}\, \s{u}
    \label{eq:geneig}
\end{equation}
where
\begin{equation} 
  \ds{L}=\left(\begin{array}{cc}
     L_1  & M  \\
     M  & L_2 
  \end{array} \right)
\end{equation}
and
\begin{equation} 
\ds{C}= 
 \left(
  \begin{array}{cc}
    {C_1}  & 0  \\
     0  & {C_2}
  \end{array} \right).
\end{equation}
The generalized eigenvalue problem has two eigenvalues, $\kappa^{\perp}_\pm$, and two current modes, $\s{u}_{\pm}$. These modes exhibit equidirected and counter-directed currents, which are called {\it Helmholtz} and {\it anti-Helmholtz} modes, respectively. They are orthogonal according to the weighted scalar product $\s{u}_\pm^\intercal \ds{C}^{-1} \s{u}_\mp=0$ and are normalized as in the following
\begin{equation}
  \varepsilon_0 l_c \, \s{u}^{\intercal}_\pm \ds{C}^{-1}\s{u}_\pm = 1.
\end{equation}
Since the magnetic energy of each current mode is strictly definite positive, the matrix $\ds{L}$ is strictly definite positive, thus $L_1 L_2 > M^2$. The solution of the problem Eq. \ref{eq:Loops2external} in terms of the current modes is 
\begin{equation}
    \s{I} = i \varepsilon_0 \omega \chi l_c \sum_{h = \pm} \frac{ \s{u}_h^\intercal \underline{\mathcal{E}} }{1 - x^2 \chi / \kappa^\perp_h } \s{u}_{h},
    \label{eq:sol2}
\end{equation}
where $\underline{\mathcal{E}}_{ext} = \left( \mathcal{E}_{1,\text{ext}}, \mathcal{E}_{2,\text{ext}} \right)^\intercal$.

\begin{figure*}[ht!]
\centering
\includegraphics[width=0.8\textwidth]{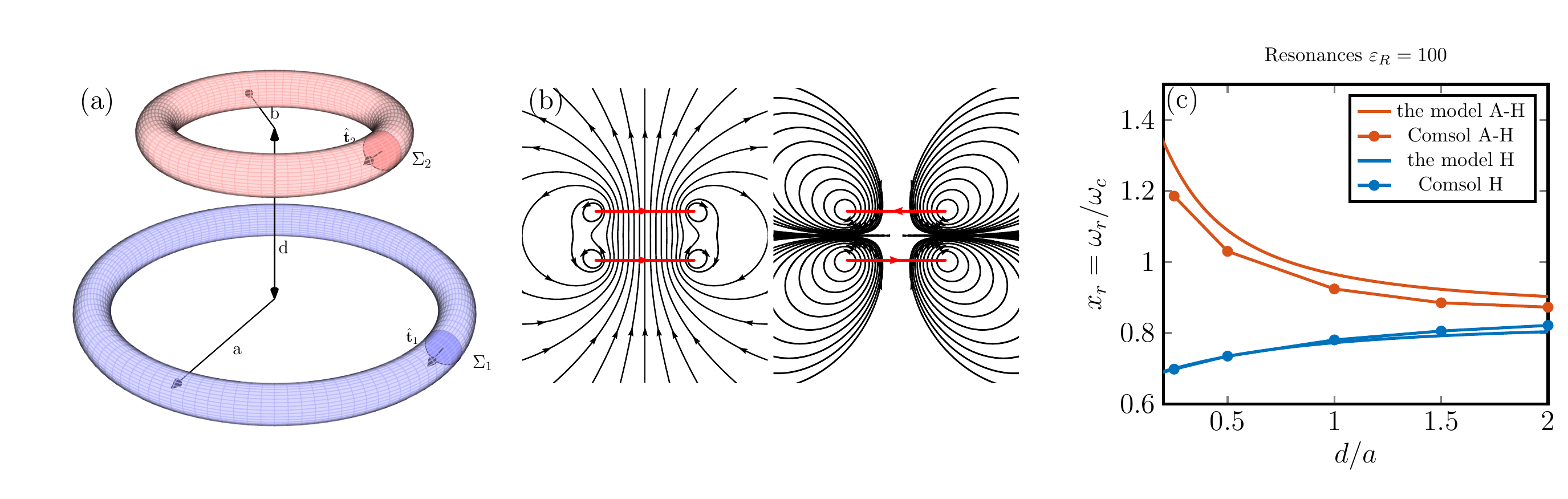}
\caption{(a) Two loops with major radii $a$ and $b$, distance $d$, sections $\Sigma_1$ and $\Sigma_2$. (b) magnetic field lines of Helmholtz (H) and anti-Helmholtz (A-H) current modes for $a=b$ and $d=a/2$. Comparison between the results obtained by the proposed model and by Comsol: resonance frequency (c) normalized to $\omega_c$, $x_{r}=\omega_{r}/\omega_c$.}
\label{fig:Loop2}
\end{figure*}

\subsection{Radiative corrections}

The second order corrections to $\kappa^{\perp}_\pm$ are given by
\begin{equation}
    \kappa^{\left( 2 \right)}_\pm =  \frac{ \left(\kappa^\perp_{\pm} \right)^2}{8\pi} \s{u}^\intercal_\pm \underline{\underline{\Delta}}^{\left( 2 \right)} \s{u}_\pm,
\end{equation}
and the first non-vanishing imaginary corrections (of odd order $n_i\ge 3$) are given by 
\begin{equation}
\kappa_\pm^{\left( n_i\right)} = i (-1)^{ \left(n_i-1\right)/2} \frac{ \left(\kappa^\perp _\pm\right)^2}{4\pi n_i !}   \s{u}_\pm^\intercal \underline{\underline{\Delta}}^{\left(n_i\right)} \s{u}_\pm
   \label{eq:KappaTE_ni},
\end{equation}
where the elements of the matrix $\underline{\underline{\Delta}}^{\left(n\right)}$ are
\begin{equation}
  \left( \underline{\underline{\Delta}}^{\left(n_i\right)}\right)_{pq} =  \oint_{\tGamma_p}   \oint_{\tGamma_q} \tv_p \rp \cdot \tv_q \rpp  \left| \rbt -\rbt'\right|^{ n_i -1} \, \dl' \dl.
\end{equation}
For loops with different shape and/or sizes, the first non-vanishing imaginary corrections for both the modes is 3. If the two loops are equal the first non-vanishing imaginary corrections for the {\it anti-Helmholtz} mode is 5.

The expression of the induced polarization currents in the loop pair with the radiative corrections is obtained by substituting $\kappa_\pm^\perp $ with $ \kappa_\pm= \gammato{\pm} + \kappa^{\left(2\right)}_\pm x^2+ i\kappa^{\left(n_i\right)}_\pm x^n_i $ into expression \ref{eq:sol2}.
\subsection{Two coaxial circular loops}

We now apply these results to two coaxial circular loops with the same permittivity $\varepsilon_R$, circular cross sections, major radii $a$ and $b$, equal minor radii $r_w=0.1a$, and axial distance $d$, as sketched in Figure \ref{fig:Loop2} (a). The self-inductances $L_1$ and $L_2$ are given by \ref{eq:SelfCircLoop} and the mutual inductance is given by \cite{maxwell_treatise_1873,grover_inductance_nodate}
\begin{equation}
   M = \mu_0\sqrt{ab} \left[ \left( \frac{2}{k'}\right) K(k')-\frac{2}{k'}E(k')\right]
    \label{eq:Mutual5}
\end{equation}
where $K \left( k \right)$ and $E \left( k \right)$ are the complete elliptic integrals of the first and second kind, and
\begin{equation}
k'=\sqrt{\frac{4 a b}{(a+b)^{2}+d^{2}}}.
\end{equation}
Figure \ref{fig:Loop2} (b) shows the field lines of the magnetic field generated by the Helmholtz $\s{u}_-$ and anti-Helmholtz $\s{u}_+$ current modes for $a=b$. Figure \ref{fig:Loop2} (c) shows  the resonance frequencies of these modes normalized to $\omega_c$, $x_{r} = \omega_{r}/\omega_c$, as a function of the distance between the loops $d$, for $b=0.8a$ and  $\varepsilon_R = 100$. The red and blue full lines represent, respectively, the resonance frequencies of the Anti-Helmholtz and Helmholtz current modes. We compare them with the peak position of the scattering response evaluated with Comsol when the two loops are excited by an electric point dipole (red and blue dashed lines) located in the equatorial plane of the torus of major radius $a$, at a distance $3a$ from its center, oriented along its toroidal direction. The agreement is very good. 

We now validate the expression of the quality factor \ref{eq:RadiativeQ} against the full-width at half maximum obtained by Comsol. Table \ref{tab:Loop2q} shows the quality factor of the Helmholtz and anti-Helmholtz modes as function of the loop distance. We compare them with the inverse of the full-width at half maximum of the peak, and we found qualitative agreement. For the anti-Helmholtz mode, since for some values of the ratio $d/a$  the overall magnetic dipole moment of the system is vanishing (or nearly vanishing), both the imaginary corrections associated to $n_i=3$ and $n_i=5$ have been considered.

\begin{table}
\centering
\begin{tabular}{lcccc}
\hline 
\multicolumn{5}{c}{Helmholtz mode} \\
$d/a$ & $0.5$ &  $1$ & $1.5$ & $2$   \\
the model &$9.9$ & $7.4$ & $6.6$ & $6.7$ \\ 
Comsol    &$12$  & $8.9$ & $7.7$ & $6.8$ \\
\hline
\multicolumn{5}{c}{Anti-Helmholtz mode} \\
$d/a$ & $0.5$ &  $1$ & $1.5$ & $2$  \\
the model &$99$ & $66$ & $34$ & $20$ \\ 
Comsol  & $38$  & $37$ & $29$ & $24$ \\
\hline 
\end{tabular}
\caption{Comparison between the radiative quality obtained by the proposed model and by Comsol for $b=0.8a$, $r_w =0.1a$, $\varepsilon_R = 100$ as a function of the distance $d/a$.}
\label{tab:Loop2q}
\end{table}

\begin{figure}[ht!]
\centering
\includegraphics[width=0.65\columnwidth]{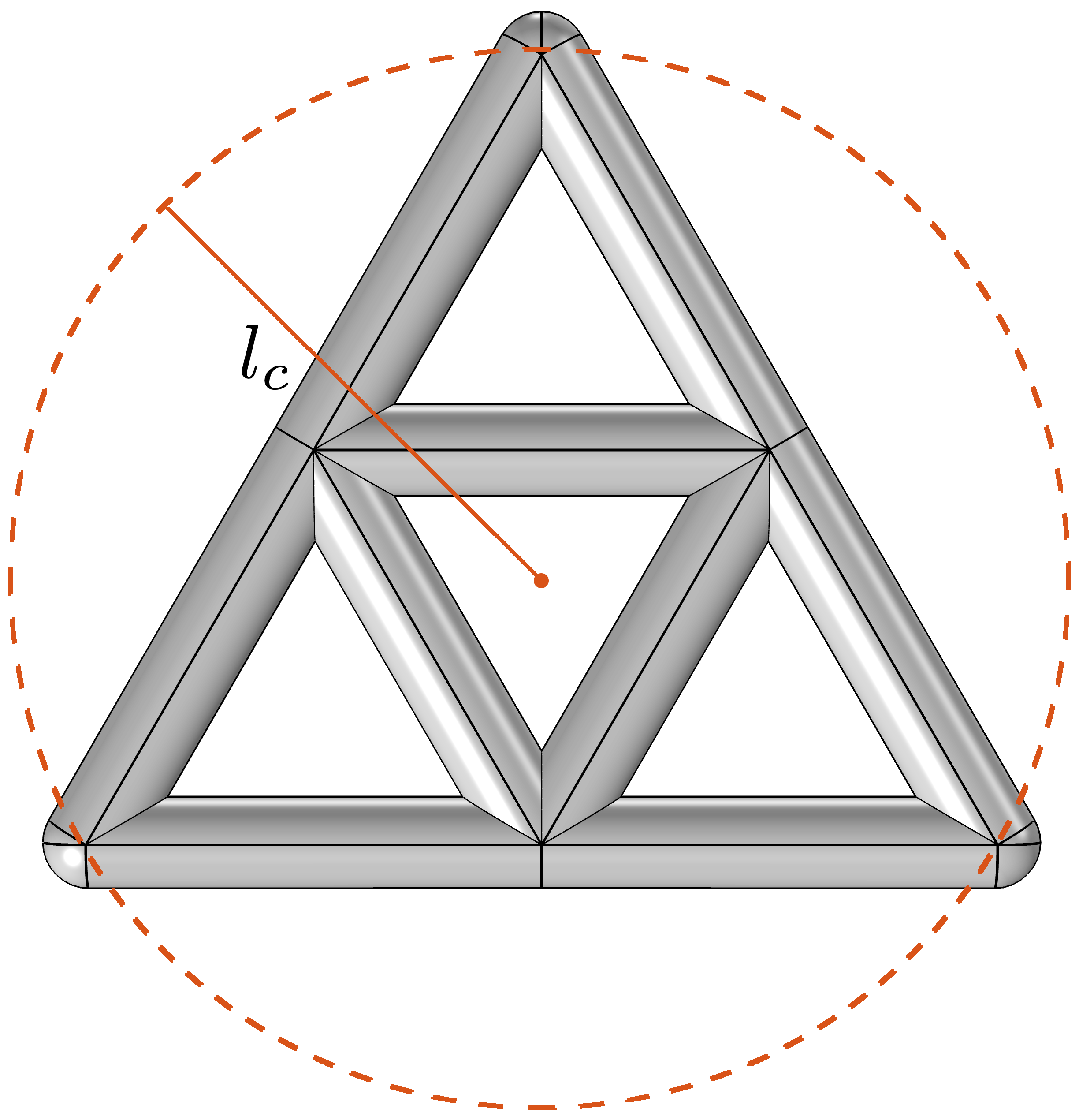}
\caption{Top view of an example of a network of dielectric wires.}
\label{fig:Sierpinski_Geom}
\end{figure}

\section{High-permittivity dielectric network}
\label{sec:circuit}
In this Section, we generalize the concepts introduced in the previous two Sections by considering an arbitrary interconnection of high-permittivity thin wires, e.g. Fig. \ref{fig:Sierpinski_Geom}. The lumped element model we propose allows to study effectively the main properties of the electromagnetic scattering from such structures: in particular, the induced polarization currents expressed in terms of the magneto-quasistatic current modes, their resonance frequencies, their radiative frequency shifts and their radiative quality factors. 

In the magneto-quasistatic limit the normal component of the polarization current density vanishes at the object's surface, hence there is not {\it leakage} of polarization current density. Therefore, at each node of the network the sum of the polarization current intensity is conserved, an analogous to the Kirchhoff's law for electric currents.
\begin{figure}[ht!]
\centering
\includegraphics[width=\columnwidth]{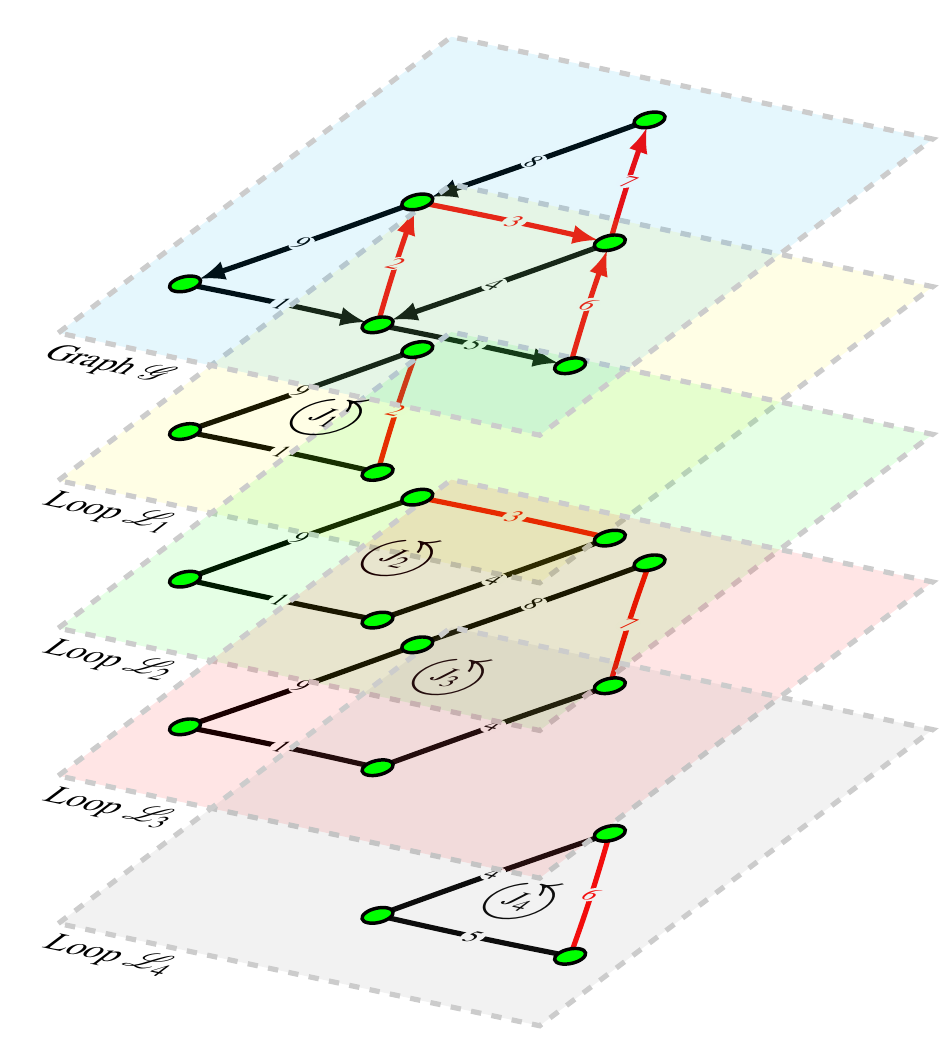}
\caption{Graph $\mathcal{G}$ associated to the high-permittivity network, where the chosen tree $\mathcal{T}$ is highlighted in red.}
\label{fig:Graph}
\end{figure}

\subsection{Network digraph}
As for electric circuits \cite{chua_linear_1987}, it is convenient to associate to the dielectric network a digraph $\mathcal{G}$, i.e. an oriented graph, with $n$ nodes and $b$ branches. We denote with $e_h$ the $h$-th branch of the digraph, $\tv_h$ its tangent unit vector, and $I_h$ the intensity of the branch polarization current. The digraph associated to the network of Fig. \ref{fig:Sierpinski_Geom} is shown on the top layer of Fig. \ref{fig:Graph}. For simplicity, we restrict our analysis to connected graphs. If the graph is not connected, one may simply combine the following approach with the one carried out in the previous section. 
A loop $\mathcal{L}_i$ of $\mathcal{G}$ is a connected sub-graph where exactly two branches are incident in each node. A tree $\mathcal{T}$ of a connected digraph $\mathcal{G}$ is a connected subgraph that contains all the nodes of $\mathcal{G}$, but no loop. For any given digraph $\mathcal{G}$, many possible choices of trees are possible. Given a connected digraph and a chosen tree, the branches of $\mathcal{T}$ are partitioned in two disjoint sets: the ones belonging to $\mathcal{T}$, called {\it twigs}, and the ones that do not belong to $\mathcal{T}$ that are called {\it links}. The {\it fundamental theorem of graphs} states that, given a connected graph with $n$ nodes and $b$ branches and a tree $\mathcal{T}$, there are $n-1$ twigs and $\it \ell=b-(n-1)$ links. Every link (e.g. the $p$-th link) together with a proper choice of twigs constitutes a unique loop, called the fundamental loop associated to the link. For instance, considering the graph shown in the top layer of Fig. \ref{fig:Graph} we can associate to every link of the particular tree $\mathcal{T}$ highlighted in red the four fundamental loops $\mathcal{L}_1, \ldots, \mathcal{L}_4$, shown in the layers below. 

A set of fundamental loops are identified through the $\it l \times b $ fundamental loop matrix $\ds{B}$ associated with the corresponding tree. The $jk$ occurrence is defined as follows: $b_{jk}=1$ if branch $k$ is in the loop $j$ and their reference directions are the same; $b_{jk}=-1$ if branch $k$ is in the loop $j$ and their reference directions are opposite; $b_{jk}=0$ if branch $k$ is not in the loop $j$.

\subsection{General lumped element model}
We now use the {\it loop analysis} to formulate the electromagnetic scattering problem from the high-permittivity network: the $b$ polarization current intensities of the network are expressed in terms of the $\it l$ fundamental loop currents. This choice  guaranties that the polarization currents satisfy the Kirchhoff's law at any node of the circuit. 

Let be $\s{I}$ the (column) vector representing the  polarization current intensities of the branches of the circuit $\left\{I_1,I_2,...,I_b\right\}$, and $\s{J}$ the column vector representing the current intensities of the links associated with the tree $\mathcal{T}$ of the circuit, $\left\{J_1,J_2,...,J_l\right\}$. The conservation of the sum of the polarization currents at the nodes of the circuit implies that
\begin{equation}
\label{eq:die1}
  \s{I} =  \ds{B}^\intercal\s{J}.
\end{equation}

Let be $\underline{\mathcal{E}}$ the column vector representing the set of induced loop voltages $\left\{ \mathcal{E}_1,\mathcal{E}_2,...,\mathcal{E}_l \right\}$ and $\underline{\mathcal{E}}_{\,ext}$ the column vector representing the set of loop external voltages $\left\{ \mathcal{E}_{ext,1},\mathcal{E}_{ext,2},...,\mathcal{E}_{ext,l} \right\}$. The constitutive relations of the dielectric thin wires give
\begin{equation}
\label{eq:die2}
  \frac{1}{\chi}\ds{B}\left(i\omega\ds{C}\right)^{-1}\s{I} = \underline{\mathcal{E}}+\underline{\mathcal{E}}_{\,ext},
\end{equation}
where $\ds{C}$ is the diagonal matrix whose elements are $C_h=(\varepsilon_0 \Sigma_h)/\Gamma_h$ with $h=1,2,..., b$, $\Sigma_h$ is the cross-section of the $h$-th wire and $\Gamma_h$ is the length.
On the other hand the Faraday-Neumann law gives 
\begin{equation}
\label{eq:die3}
  \underline{\mathcal{E}} = -i\omega \ds{L} \,\s{J}
\end{equation}
where $\ds{L}$ is the $\it l \times \l$ inductance matrix of the set of fundamental loops. By combining equations \ref{eq:die1}-\ref{eq:die3} we obtain the system of equations governing the set of the link currents associated to the tree $\mathcal{T}$,
\begin{equation}
\label{eq:die4}
 i\omega \ds{L} \,\s{J}+\frac{1}{i\omega\chi}\ds{B}\, \ds{C}^{-1} \ds{B}^\intercal \s{J}=\underline{\mathcal{E}}_{\,ext}.
\end{equation}
The current modes $\left\{ \s{u}_h\right\}$ of the network and the corresponding eigenvalues $\left\{ \kappa_h^\perp \right\}$ are solution of the generalized eigenvalue problem
\begin{equation}
    \ds{L} \, \s{u} = \frac{1}{\kappa^\perp}\frac{1}{\omega_c^2}\ds{B}\, \ds{C}^{-1} \ds{B}^\intercal\, \s{u}.
    \label{eq:geneig}
\end{equation}
The number of magneto-quasistatic current modes and of resonances of a high-permittivity dielectric network is equal to the number $l$ of links of the digraph $\mathcal{G}$ of the network. The matrices $\ds{L}$ and $\ds{B}\, \ds{C}^{-1} \ds{B}^\intercal$ are symmetric and definite positive. As a consequence, the eigenvalues $\left\{ \kappa_h^\perp \right\}$ are real and positive, and the current modes satisfy a weighted orthogonality, 
\begin{equation}
    \varepsilon_0 l_c \s{u}_h^\intercal \left(\ds{B}\, \ds{C}^{-1} \ds{B}^\intercal \right) \s{u}_k = \delta_{hk}.
\end{equation}
The solution of Eq. \ref{eq:die4} is
\begin{equation}
    \s{J} = i \varepsilon_0 \omega \chi l_c \sum_{h = 1}^\ell \frac{ \s{u}_h^\intercal \underline{\mathcal{E}}_{ext} }{1 - x^2 \chi / \kappa^\perp_h } \s{u}_{h}. 
    \label{eq:cur3}
\end{equation}

\subsection{Radiative corrections}
The expression of the second order (real) radiative correction for the $h-th$ mode is
\begin{equation}
\kappa_h^{\left( 2 \right)} =  \frac{ \left(\gammato{h}\right)^2}{8\pi}   \s{u}_h^\intercal \underline{\underline{\Delta}}^{\left(2\right)} \s{u}_h,
   \label{eq:Kappa2TE} 
\end{equation}
and the expression of the lowest  order $n_i$ imaginary correction (which is an odd number $n_i \ge 3$) is
\begin{equation}
\kappa_h^{\left( n_i \right)} = i \left( - 1 \right)^{ \left( n_i - 1 \right)/2} \frac{ \left(\gammato{h}\right)^2}{4\pi \left(n_i\right)!}  \, \s{u}_h^\intercal \, \ds{B} \, \underline{\underline{\Delta}}^{\left(n_i\right)} \, \ds{B}^\intercal \s{u}_h
   \label{eq:KappaTE_ni} 
\end{equation}
where the elements of the matrix $\underline{\underline{\Delta}}^{\left(n\right)}$ are
\begin{equation}
    \left( \underline{\underline{\Delta}}^{\left( n \right)} \right)_{ij} = \oint_{e_i}  \tv_i \rp\cdot \oint_{e_j}  \tv_j \rpp {\left| \rbt -\rbt'\right|^{n-1}}  \dl' \dl.
\end{equation}

The expression of the induced polarization currents in the links of the dielectric network taking into account the radiative corrections is obtained by substituting $\kappa_h^\perp $ with $ \kappa_h= \gammato{h} + \kappa^{\left(2\right)}_h x^2+ i\kappa^{\left(n_i\right)}_h x^n_i$ for $h=1,\ell$ into expression \ref{eq:cur3}.

\subsection{Partial inductances}
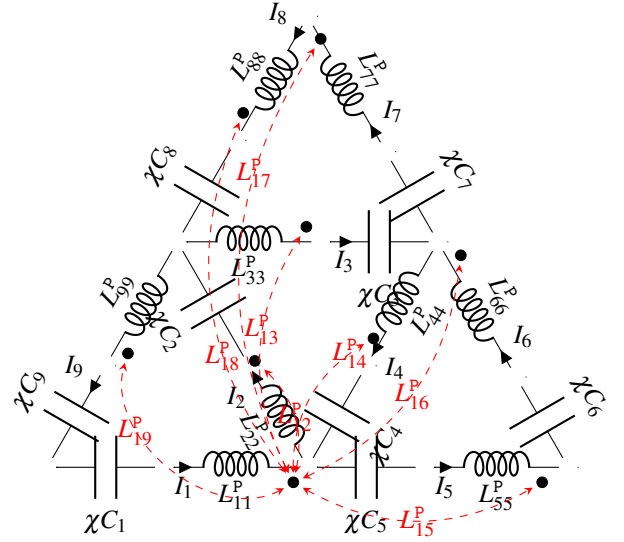
\begin{figure}
\centering
\begin{circuitikz}
\draw [fill=black] (-4*0.8660,-4*0.5000)node(a){};
\draw [fill=black] (-4*0.4330,+4*0.2500)node(b){};

\draw [fill=black] (0,-4*0.5000)node(c){};
\draw [fill=black] (+4*0.8660,-4*0.5000)node(d){};
\draw [fill=black] (+4*0.4330,+4*0.2500)node(e){};
\draw [fill=black] (0,+4)node(f){};
\node (ab) at ($(a)!0.45!(b)$) {};
\node (ac) at ($(a)!0.45!(c)$) {};
\node (bc) at ($(b)!0.55!(c)$) {};
\node (ce) at ($(c)!0.45!(e)$) {};
\node (be) at ($(b)!0.55!(e)$) {};
\node (cd) at ($(c)!0.45!(d)$) {};
\node (de) at ($(d)!0.4!(e)$) {};
\node (bf) at ($(b)!0.45!(f)$) {};
\node (ef) at ($(e)!0.4!(f)$) {};

\draw (b) to [L,l_=$L_{99}^\mathtt{P}$] (ab);
\draw (ab) to [C,l_=$\chi C_9$,i>_=$I_9$] (a);
\draw [fill=black] ($(b)!0.9!(ab)+(0.2,0)$)node(n9){} circle (2pt);

\draw (a) to [C,l_=$\chi C_1$]   (ac);
\draw (ac) to [L,l_=$L_{11}^\mathtt{P}$,i>_=$I_1$] (c);
\draw [fill=black] ($(ac)!0.9!(c)+(0,-0.2)$)node(n1){} circle (2pt);

\draw (b) to  [C,l_=$\chi C_2$] (bc);
\draw (bc) to [L,l_=$L_{22}^\mathtt{P}$,i<_=$I_2$]  (c);
\draw [fill=black] ($(c)!0.9!(bc)+(0,0.2)$)node(n2){} circle (2pt);

\draw (c) to [C,l_=$\chi C_4$]  (ce); 
\draw (ce) to [L,l_=$L_{44}^\mathtt{P}$,i<_=$I_4$]  (e);
\draw [fill=black] ($(ce)!0.1!(e)+(0,0.2)$)node(n4){} circle (2pt);

\draw (b) to [L,l_=$L_{33}^\mathtt{P}$] (be);
\draw (be) to [C,l_=$\chi C_3$,i>_=$I_3$]  (e); 
\draw [fill=black] ($(b)!0.9!(be)+(0,0.2)$)node(n3){} circle (2pt);

\draw (c) to [C,l_=$\chi C_5$]  (cd);
\draw (cd) to [L,l_=$L_{55}^\mathtt{P}$,i>_=$I_5$] (d); 
\draw [fill=black] ($(c)!0.9!(d)+(0,-0.2)$)node(n5){} circle (2pt);

\draw (d) to [C,l_=$\chi C_6$] (de);
\draw (de) to  [L,l_=$L_{66}^\mathtt{P}$,i>_=$I_6$] (e); 
\draw [fill=black] ($(de)!0.9!(e)+(0.2,0)$)node(n6){} circle (2pt);

\draw (f) to [L,l_=$L_{88}^\mathtt{P}$,i>_=$I_8$] (bf);
\draw (bf) to [C,l_=$\chi C_8$]  (b); 
\draw [fill=black] ($(f)!0.9!(bf)+(0,0.2)$)node(n8){} circle (2pt);

\draw (e) to [C,l_=$\chi C_7$] (ef);
\draw (ef) to [L,l_=$L_{77}^\mathtt{P}$,i>_=$I_7$]   (f); 
\draw [fill=black] ($(e)!0.9!(f)+(0,0)$)node(n7){} circle (2pt);

\draw [<->,>=stealth, dashed, red] (n1)  to [bend left]  node[pos=0.60] {$L_{13}^\mathtt{P}$} (n3);
\draw [<->,>=stealth, dashed, red] (n1)  to [bend right] node[pos=0.5]  {$L_{12}^\mathtt{P}$} (n2);
\draw [<->,>=stealth, dashed, red] (n1)  to [bend left]  node[pos=0.9] {$L_{14}^\mathtt{P}$} (n4);
\draw [<->,>=stealth, dashed, red] (n1)  to [bend right] node[pos=0.5]  {$L_{15}^\mathtt{P}$} (n5);
\draw [<->,>=stealth, dashed, red] (n1)  to [bend right, bend angle=70] node[pos=0.5]  {$L_{16}^\mathtt{P}$} (n6);
\draw [<->,>=stealth, dashed, red] (n1)  to [bend left, bend angle=0]  node[pos=0.7] {$L_{17}^\mathtt{P}$} (n7);
\draw [<->,>=stealth, dashed, red] (n1)  to [bend left,  bend angle=85]  node[pos=0.35]  {$L_{18}^\mathtt{P}$} (n8);
\draw [<->,>=stealth,bend angle=70, dashed, red] (n1)  to [bend left] node[pos=0.7] {$L_{19}^\mathtt{P}$} (n9);
\end{circuitikz}
\caption{Lumped element circuit of capacitances and partial inductances. The self partial inductances $L_{qq}^\texttt{P}$ of each branch of the digraph are shown. The mutual partial inductances between the first branch and any other branch are shown in red. The dotted convention commonly employed in magnetically coupled circuits is used in this case.}
\label{fig:Sierpinski_Circuit}
\end{figure}
The direct calculation of self- and mutual- inductance of fundamental loops may not be the most efficient method to assembly the matrix $\ds{L}$, since different loops may share several branches, resulting in redundant hence inefficient computations. It is instead convenient, as for electric circuits \cite{grover_inductance_nodate,paul_inductance_2011}, to preliminary assemble the partial loop inductances matrix $\ds{L}^\texttt{P}$. Its $ij$- occurrence is the partial inductance $L_{ij}^{\text{P}}$ between the branches $e_i$ and $e_j$. It is defined as the ratio between the magnetic flux produced by the density current flowing in the branch $e_i$, through the surface between the second branch $e_j$ and infinity, and the current of the branch $e_i$, namely:
\begin{equation}
    L_{ij}^{\texttt{P}} = \frac{\mu_0}{4\pi} \oint_{e_i} \oint_{e_j} \frac{ \tv_i \left( \rb \right) \cdot \tv_j \left( \rb' \right)}{\left| {\bf r} - {\bf r}' \right|} d\tilde{l}' d\tilde{l}.
\end{equation}
 There are  $b\left(b+1\right)/2$ independent partial inductance $L_{ij}^{\text{P}}$, because $L_{ij}^{\texttt{P}} =L_{ji}^{\texttt{P}} $ by reciprocity. The loop inductance matrix $\ds{L}$ is given by
\begin{equation}
   \ds{L} = \ds{B} \, \ds{L}^\text{P} \, \ds{B}^\intercal
\end{equation}

If the circuit is composed by an arbitrary interconnection of {\it straight} wires laying on the same plane, then we need formulas for calculating: i) self-partial inductance of a straight wire; ii) the mutual partial inductance between wires at an angle to each other which includes as a limit case the mutual partial inductance between parallel wires, and the mutual partial inductance of wires meeting in a point (given by G.A. Campbell \cite{campbell_mutual_1915}). If the circuit is not planar, then also the mutual partial inductance between skewed and displacement wires first derived by F. F. Martens and G. A. Campbell \cite{martens_uber_1909,campbell_mutual_1915,grover_inductance_nodate,paul_inductance_2011} are needed. All these formulas have been analytically derived a century ago, and are reported in the Appendix.
In Fig. \ref{fig:Sierpinski_Circuit} we show the lumped circuit for the dielectric network shown in Fig. \ref{fig:Sierpinski_Geom}, where we illustrate the self partial inductances $L_{qq}^\texttt{P}$, and for sake of clarity the mutual partial inductance $L_{1q}^\texttt{P}$. 

\subsection{Sierpinski triangle}

Let us analyze the dielectric network shown in Fig. \ref{fig:Graph}, constituted by $b=9$ high-permittivity thin wires of equal length $l_w$, with circular cross section with radius $r_w=0.1 l_w$. The wires are interconnected accordingly to a Sierpinski triangle. The minimum circle circumscribing the network is chosen as characteristic length $l_c$. 

\begin{figure}
\centering
\includegraphics[width=\columnwidth]{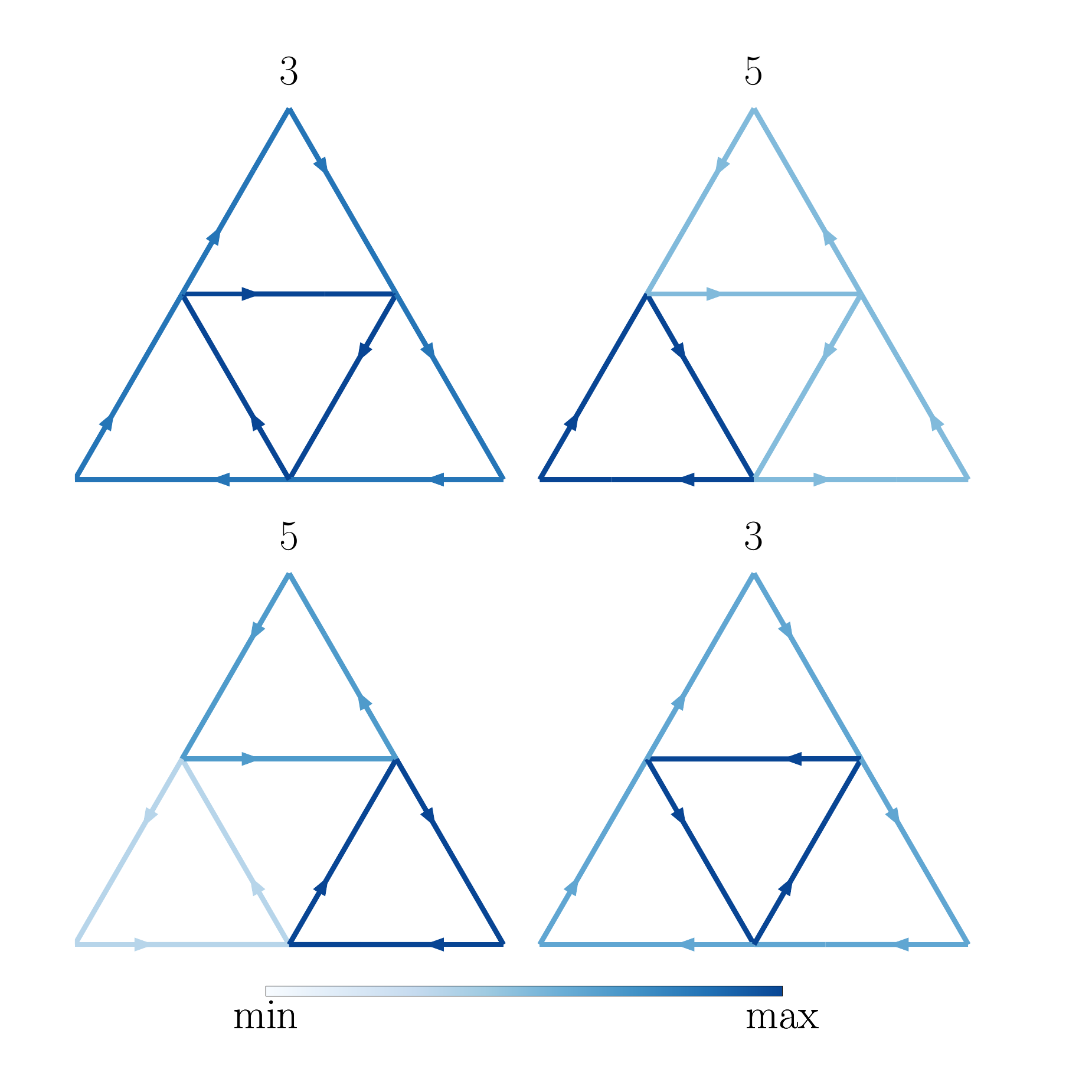}
\caption{Current modes of the dielectric network arranged accordingly to a Sierpinsky triangle with generation number 1. Each wire has length $l_w$ and radius $r_w = 0.1l_w$. The modes are arranged with a lexicographic order which follows the corresponding eigenvalue. Above each mode is the order $n_i$ of the first non-vanishing imaginary correction.}
\label{fig:Sierpinski_n1}
\end{figure} 

The graph of this network is shown in the top layer of Fig. \ref{fig:Graph}, where the twigs and links, associated to a chosen tree, are highlighted in red and black, respectively. The fundamental loops associated to each {\it link} are shown in the layers below. The lumped element circuit is shown in Fig. \ref{fig:Sierpinski_Circuit}. The $45$ independent partial inductances are firstly evaluated, then the $10$ independent elements of the  $4 \times 4$ symmetric inductance matrix $\underline{\underline{\text{L}}}$ are computed. In this simple case, the calculation of the eigenvalues $\kappa^{\left(2\right)}_h$ can be carried out with pen and paper, returning the four values listed in Tab. \ref{tab:Sierpinski_kappa}. The four current modes are shown in Fig. \ref{fig:Sierpinski_n1} (the second and the third mode are degenerate). These modes correspond to the magneto-quasistatic current density modes reported in \ref{fig:Sierpinski_n1}. Then, the  matrices  $\underline{\underline{\Delta}}^{\left(2\right)}$ and $\underline{\underline{\Delta}}^{\left(n_i\right)}$are assembled, where $n_i=3$ for the first and fourth mode, and $n_i=5$ for the second and third mode. The radiation corrections $\kappa^{\left(2\right)}_h$ and $\kappa^{\left(n_i\right)}_h$ are given in Tab.\ref{tab:Sierpinski_kappa}.

\begin{table}
\centering
\begin{tabular}{lcccc}
\hline 
$\kappa^{\perp}_h$ & 134 & 181 & 181 & 379  \\
$\kappa^{\left(2\right)}_h$ &   -20.1 & -6.7 & -6.7 & -26.0 \\
$n_i$ & 3 & 5 & 5 & 3 \\
$\kappa^{\left(n_i\right)}_h$ &  9.54 & 0.33 &  0.33 &  11.2 \\
\hline 
\end{tabular}
\caption{Eigenvalues $\kappa^{\perp}_h$, second order (real) radiative correction $\kappa^{\left(2\right)}_h$, imaginary correction $\kappa^{\left(n_i\right)}_h$ of the lowest order $n_i$.}
\label{tab:Sierpinski_kappa}
\end{table}

\begin{table}
\centering
\begin{tabular}{lccccccc}
\hline 
\textbf{$\varepsilon_r=100$}  &
\textbf{$x_{r,1}$} & \textbf{$x_{r,2}$}      & \textbf{$x_{r,3}$}                       & \textbf{$x_{r,4}$}   & \textbf{$Q_1$} & \textbf{$Q_2=Q_3$} & \textbf{$Q_4$}   \\
the model                                         & $1.06$ & $1.30$      & $1.30$& $1.74$  & 11.7  & 244 & 6.5 \\ 
Comsol & $1.04$ & $1.26$ & $1.26$ & $-$ & $31$ & $280$ & $-$\\
\hline \textbf{$\varepsilon_r=15.45$} &
\textbf{$x_{r,1}$} & \textbf{$x_{r,2}$}      & \textbf{$x_{r,3}$}                       & \textbf{$x_{r,4}$}   & \textbf{$Q_1$}  & \textbf{$Q_2=Q_3$}  & \textbf{$Q_4$}          \\ 
the model                                              & $1.94$ & $2.86$      & $2.86$ & $3.02$  & 1.9 & 23 & 1.2 \\ 
Comsol & $2.29$ & $2.89$ & $2.89$ & $-$ & $3.9$ & $8.3$ & $-$\\
\hline
\end{tabular}
\caption{Normalized resonance frequencies $x_{h}=\omega_{h}/\omega_c$ and quality factors of the four modes of the dielectric network of Fig. \ref{fig:Sierpinski_Geom} for $\varepsilon_R = 100$ and $\varepsilon_R = 15.54$ }
\label{tab:Sierpinski_Res}
\end{table}

Table \ref{tab:Sierpinski_Res} gives the normalized resonance frequencies and the quality factors of the four current modes of the Sierpinsky network for two different values of the permittivity, $\varepsilon_R=100$ and $\varepsilon_R=15.45$.
For $\varepsilon_R=100$ the first (fundamental) resonance is located at $x_1=1.04$: for $l_c=10cm$ it is $\omega_1 = 3.14 GHz$. The scattering peak positions and the quality factors of the four current modes have been also estimated by using Comsol Multiphysics. The dielectric network shown in Fig. \ref{fig:Sierpinski_Geom} has been excited by an electric dipole, laying on its equatorial plane, located at $3l_w$ on the left of its center, and oriented along the vertical in-plane direction. The quality factor is estimated as the inverse of the full-width at half maximum. The lumped circuit model exhibits a good accuracy in locating the resonances, while returning the order of magnitude of the quality factors. The comparison is repeated for $\varepsilon=15.45$. The first (fundamental) resonance is located at $x_1=1.94$, thus for $l_c=250$nm the resonance wavelength is located within the visible spectral range, $\lambda_1 =810$nm. The relative error in the predicted resonance position (compared to its Comsol counterpart) worsen on average with respect to the previous scenario, but it remains below the $15 \%$. For both the investigated permittivities, in the Comsol simulation, the scattering peak associated to the fourth mode is not clearly identifiable, due to its low quality factor. For this reason, the position of its resonance peak and its Q-factor estimate are omitted in Tab. \ref{tab:Sierpinski_Res}.

\begin{figure}
\centering
\includegraphics[width=\columnwidth]{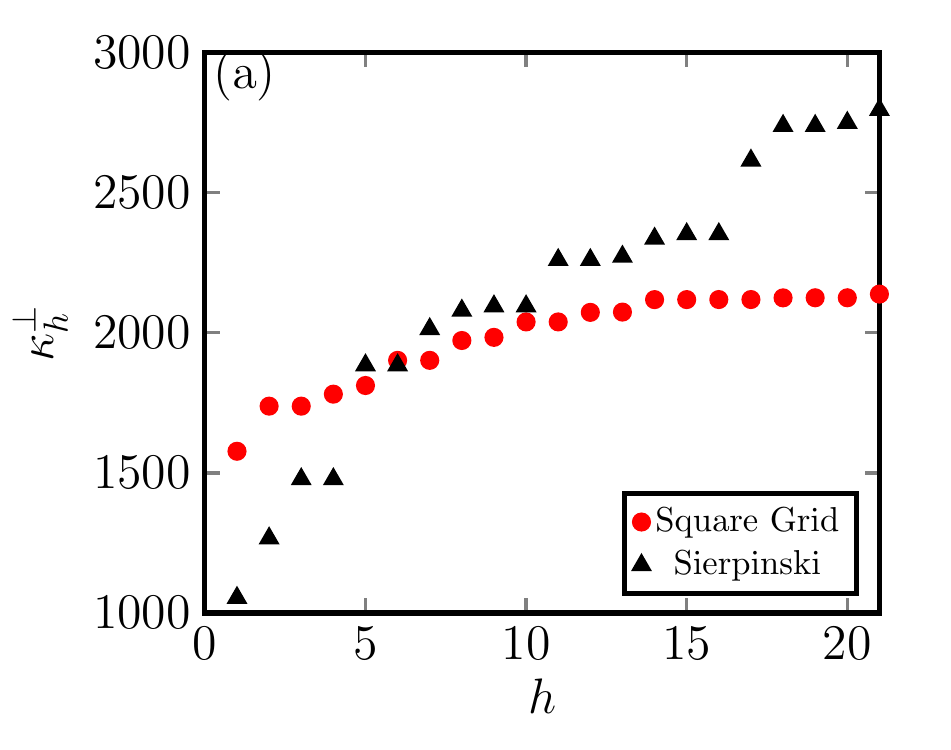}
\caption{First 21 eigenvalues $\kappa^{\perp}_h$ of a Sierpinski triangle (generation number 3) with $b=81,n=42,\ell=37$, and first 21 eigenvalues $\kappa^{\perp}_h$ of a $7\times7$ square grid with  $b=112,n=64,\ell=49$.}
\label{fig:Comparsion}
\end{figure}

\begin{figure*}
\centering
\includegraphics[width=\textwidth]{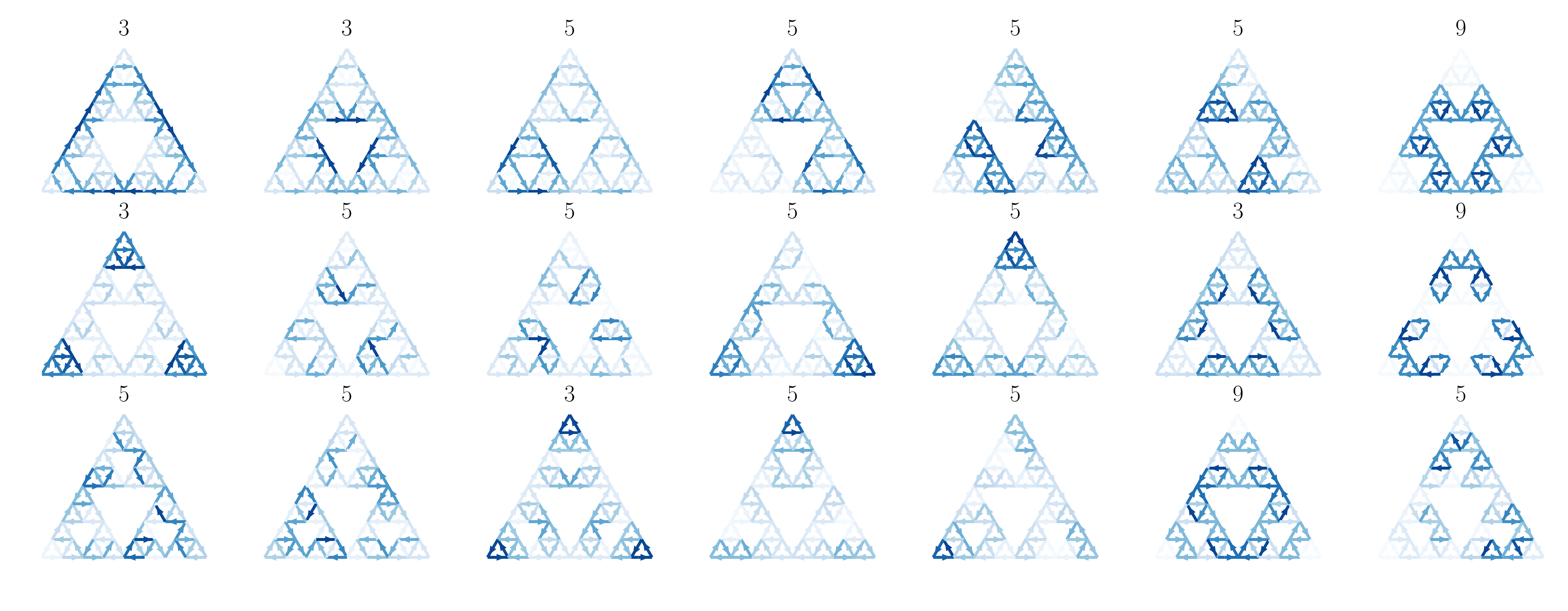}
\caption{First $21$ MQS modes of a high permittivity circuit made of $b=81$ thin wires interconnected accordingly to a Sierpinski triangle. Each wire has length $a$ and radius $r_w = 0.1 a$. The modes are lexicographically ordered in terms of increasing MQS eigenvalues. Above each mode is the order $n_i$ of the first non-vanishing imaginary correction.}
\label{fig:Sierpinski_n3}
\end{figure*}

\begin{figure*}
\centering
\includegraphics[width=\textwidth]{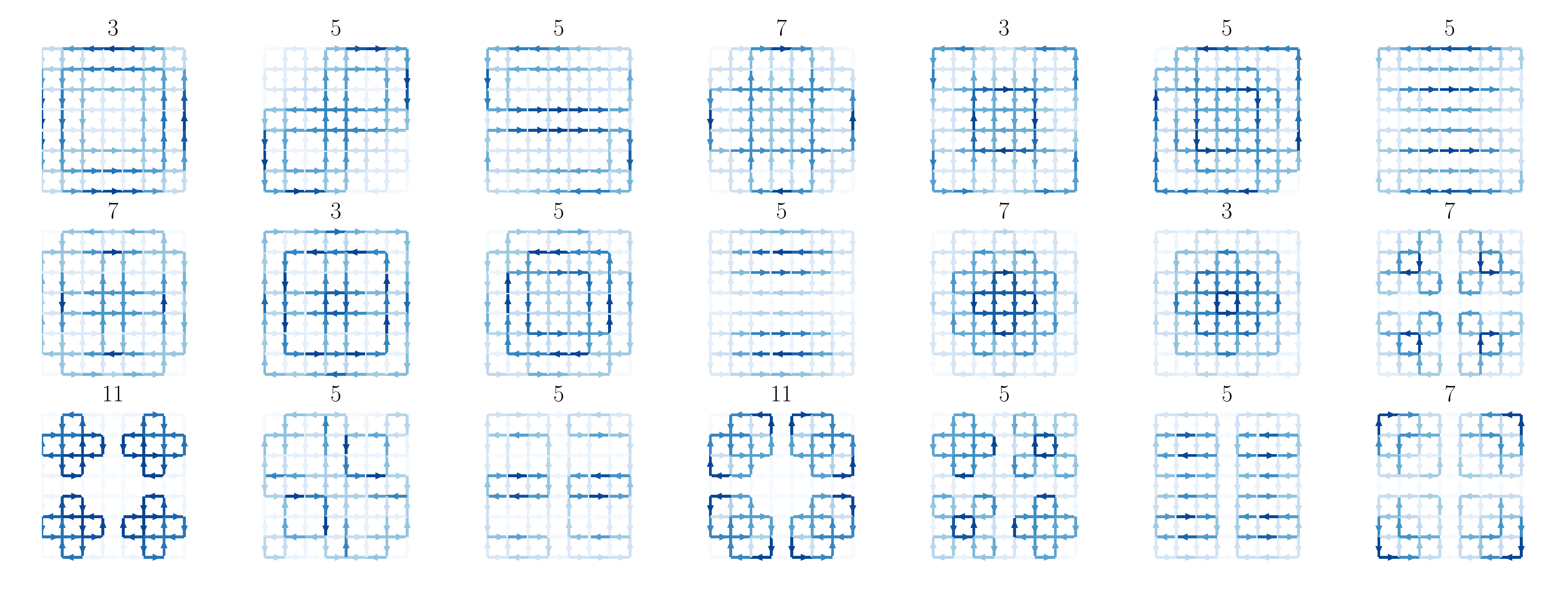}
\caption{First 21 MQS modes of a high permittivity network made of $b=112$ thin wires arranged accordingly to a square grid. Each wire has length $a$ and radius $r_w = 0.1 a$. The modes are arranged with a lexicographic order which follows the increasing MQS eigenvalues. Above each mode is the order $n_i$ of the first non-vanishing imaginary correction.}
\label{fig:Square}
\end{figure*}

We now illustrate how the analysis of complex high-permittivity networks may benefit from the use of a lumped-elements approach. We investigate two networks, the first one is composed of $b=81$ thin wires interconnected accordingly to a Sierpinski triangle with generation number $3$; the second one is instead composed of $b=112$ thin wires, interconnected accordingly to a square grid. The first (smallest) $21$ eigenvalues of both the networks are compared in Fig. \ref{fig:Comparsion}. The corresponding current modes are shown in Figs. \ref{fig:Sierpinski_n3} and \ref{fig:Square}. The number reported above each mode is the order $n_i$ of the first non-vanishing imaginary radiative correction. This number returns the power dependence of the quality factor on the size-parameter, accordingly to Eq. \ref{eq:RadiativeQ}, which is related to the multipolar components of the mode. 
\section{Conclusions}



We investigated the electromagnetic scattering from {\it high-permittivity dielectric networks}. They are interconnections of high-permittivity dielectric thin wires. If the overall size of the network is smaller than (or at most equal to) the operating wavelength the dielectric network can be modeled as a lumped circuit constituted by capacitances, self- and mutual- loop inductances. The resonant modes are equal in number to the links of the network's digraph and are related to the spectrum of the loop-inductance matrix. Closed form expressions are given for the frequency shifts and quality factors due to the radiation.  The inductance matrix can be assembled from the {\it partial} self- and mutual- inductance of the constituent wires, transplanting to the electromagnetic scattering theory formulas introduced a century ago for inductive network.  For networks with sizes smaller than the incident wavelength, the error in locating the resonance is acceptable, below $15 \%$ in the investigated numerical experiments. This fact promotes the proposed model as a fast computational tool for preliminary analysis, that could be later refined by more accurate tools that are usually associated with much higher computational burden. This manuscript may also stimulate the grafting of several other ideas and methods from the electric and electronic circuit onto design of high-index resonators in both the microwaves and visible spectral range. This approach together with the one proposed in Ref. \cite{forestiere_electromagnetic_2019} represent the first steps toward the derivation of a full-wave treatment of complex networks of wires.

\appendix

\section{Partial Inductances Calculation}
\label{app:partial}

\begin{figure}
    \centering
    \includegraphics[width=2cm]{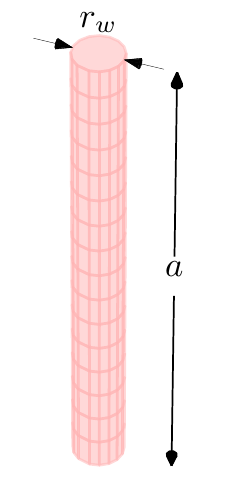}
    \caption{Self partial inductance of a wire of length $a$ and circular cross section of radius $r_w$.}
    \label{fig:SingleWire}
\end{figure}

\begin{figure}
    \centering
    \includegraphics[width=6cm]{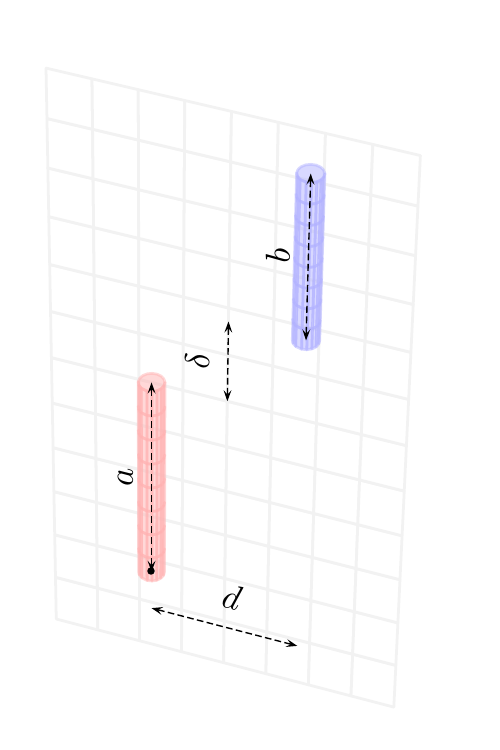}
    \caption{Mutual partial inductance of two wires of lengths $a$ and $b$, equal circular cross section of radius $r_w$, at distance $d$ and offset $z$.}
    \label{fig:TwoWires}
\end{figure}

\begin{figure}
    \centering
    \includegraphics[width=6cm]{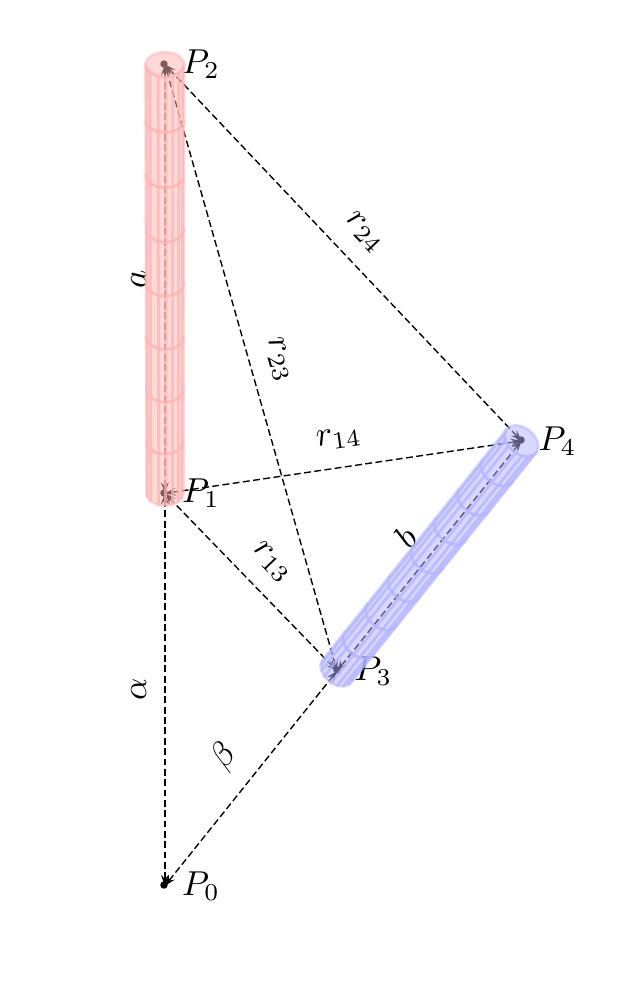}
     \caption{Mutual partial inductance of two wires of lengths $a$ and $b$, at an angle to each other.}
    \label{fig:TwoWiresAngle}
\end{figure}

\subsection{Self partial inductance of a straight wire}
The self partial inductance $L_p$ of the wire of Fig. \ref{fig:SingleWire} with radius $r_w$ and length $a$ is
\begin{multline}
    L^\texttt{P} = \frac{a \mu_{0}}{2 \pi} \left[\ln \left(\frac{a}{r_{\mathrm{w}}}+\sqrt{\left(\frac{a}{r_{\mathrm{w}}}\right)^{2}+1}\right) \right. \\ \left. -\sqrt{1+\left(\frac{r_{\mathrm{w}}}{a}\right)^{2}}+\frac{r_{\mathrm{w}}}{a}\right] + \frac{\mu_{0}}{8\pi}a.
\end{multline}

\subsection{Mutual partial inductance between two unequal parallel wires that are offset}
The mutual partial inductance $M_p$ between the two wires shown in Fig. \ref{fig:TwoWires} of negligible cross section, length $a$ and $b$. at a distance $d$, with an offset $\delta$, is given by \cite{eccles_wireless_1918,grover_inductance_nodate}:
\begin{multline}
    {M^\texttt{P}} = \frac{\mu_{0}}{4 \pi}\left[z_{2} \sinh ^{-1} \frac{z_{2}}{d}\right. \\-z_{1} \sinh ^{-1} \frac{z_{1}}{d}  -\left(z_{2}-a\right) \sinh ^{-1} \frac{z_{2}-a}{d} \\ +\left(z_{1}-a\right) \sinh ^{-1} \frac{z_{1}-a}{d}-\sqrt{z_{2}^{2}+d^{2}}+\sqrt{z_{1}^{2}+d^{2}} \\ \left.+\sqrt{\left(z_{2}-a\right)^{2}+d^{2}}-\sqrt{\left(z_{1}-a\right)^{2}+d^{2}}\right]
\end{multline}
where $z_2=a+b+\delta$ and $z_1=a+\delta$.

\subsection{Mutual partial inductance between wires at an angle to each other}
Let us consider the two wires of Fig. \ref{fig:TwoWiresAngle}, of length $a$ and $b$, of negligible cross section (filaments). The wires are coplanar, forming an angle $\theta$ to each other \cite{campbell_mutual_1915,grover_inductance_nodate,paul_inductance_2011}. Their mutual inductance has the following expression:
\begin{multline}
     M^\texttt{P}  = \frac{\mu_{0}}{4 \pi}\left[\left(\beta+b\right) \ln \frac{r_{24}+r_{14}+a}{r_{24}+r_{14}-a} - \beta \ln \frac{r_{23}+r_{13}+a}{r_{23}+r_{13}-a}  \right. \\ + \left. \left(a+\alpha\right) \ln \frac{r_{24}+r_{23}+b}{r_{24}+r_{23}-b}-\alpha \ln \frac{r_{14}+r_{13}+b}{r_{14}+r_{13}-b}\right]
\end{multline}
if the two wires are touching in $P_0$ then the above equation reduces to \cite{campbell_mutual_1915}:
\begin{equation}
  M^\texttt{P}  = \frac{\mu_{0}}{4 \pi} \cos \theta\left(a \ln \frac{r_{24}+a+b}{r_{24}+a-b} + b \ln \frac{r_{24}+a+b}{r_{24}+b-a}\right).
\end{equation}

\end{document}